\documentclass[prl,aps,twocolumn,groupedaddress,showpacs]{revtex4}

\usepackage{bm}
\usepackage{graphicx}
\usepackage{color}

\begin{document}

\title{Observation of Two-Dimensional Bulk Electronic States in a Superconducting Topological Insulator Heterostructure Cu$_x$(PbSe)$_5$(Bi$_2$Se$_3$)$_6$: Implications for Unconventional Superconductivity}

\author{K. Nakayama,$^1$ H. Kimizuka,$^1$ Y. Tanaka,$^1$ T. Sato,$^1$ S. Souma,$^2$ T. Takahashi,$^{1,2}$ Satoshi Sasaki,$^{3,\ast}$ Kouji Segawa,$^{3,\S}$ and Yoichi Ando$^3$}

\affiliation{$^1$Department of Physics, Tohoku University, Sendai 980-8578, Japan\\
$^2$WPI Research Center, Advanced Institute for Materials Research, Tohoku University, Sendai 980-8577, Japan\\
$^3$Institute of Scientific and Industrial Research, Osaka University, Ibaraki, Osaka 567-0047, Japan
}

\date{\today}

\begin{abstract}
We have performed angle-resolved photoemission spectroscopy (ARPES) on Cu$_x$(PbSe)$_5$(Bi$_2$Se$_3$)$_6$ (CPSBS; $x$ = 1.47), a superconductor derived from a topological insulator heterostructure, to elucidate the electronic states relevant to the occurrence of possible unconventional superconductivity. Upon Cu intercalation into the parent compound (PbSe)$_5$(Bi$_2$Se$_3$)$_6$, we observed a distinct energy shift of the bulk conduction band due to electron doping. Photon-energy dependent ARPES measurements of CPSBS revealed that the observed bulk band forms a cylindrical electronlike Fermi surface at the Brillouin-zone center. The two-dimensional nature of the bulk electronic states suggests the occurrence of odd-parity $E_u$ pairing or even-parity $d$-wave pairing, both of which may provide a platform of Majorana bound states in the superconducting state.
\end{abstract}


\maketitle
Topological superconductors (TSCs) manifest a novel quantum state of mater in which the nontrivial topology of the bulk state leads to the emergence of gapless Andreev bound states often consisting of Majorana fermions \cite{HasanReview, ZhangReview, AndoReview, AliceaReview, BeenakkerReview}. Owing to the exotic characteristics of Majorana fermions such as particle-antiparticle symmetry and potential use for fault-tolerant quantum computations, the search for TSCs is one of the most attractive and emergent topics in condensed matter physics. It has been theoretically proposed that topological insulators (TIs), which are characterized by the gapless topological edge or surface states originating from a band inversion due to strong spin-orbit coupling, can provide a platform to realize topological superconductivity when enough carriers are introduced \cite{FuTSC, MSatoPRB}. A first example of the superconductor derived from TIs is Cu$_x$Bi$_2$Se$_3$ \cite{Hor, KrienerPRL}, in which point-contact spectroscopy experiments observed a pronounced zero-bias conductance peak indicative of unconventional surface Andreev bound states \cite{SasakiCuBi2Se3}. Subsequent experiment in a doped topological crystalline insulator (TCI) Sn$_{1-x}$In$_x$Te \cite{SasakiInSnTe} has also reported a similar Andreev bound states. These results suggest that the carrier doping into the TIs and TCIs is an effective strategy to search for three-dimensional (3D) TSC, while the electronic states (in particular, the superconducting pairing symmetry) and their relationship to possible TSC nature are still under intensive debate \cite{KrienerPRL, SasakiCuBi2Se3, SasakiInSnTe, KanigelZBCP, CuBi2Se3STM, PengAndreev, HasanCuBi2Se3, KanigelARPES, KrienerPRB, CuBi2Se3Bc2, SatoInSnTe, NovakInSnTe}.

Very recently, it has been suggested that a new TI-based superconductor Cu$_x$(PbSe)$_5$(Bi$_2$Se$_3$)$_6$ (called CPSBS here) is an intriguing candidate of TSC \cite{SasakiCPSBS}. The parent compound is the $m = 2$ phase of lead-based homologous series (PbSe)$_5$(Bi$_2$Se$_3$)$_{3m}$ (called PSBS), in which Bi$_2$Se$_3$ with the thickness of $m$ quintuple layers (QLs) alternates with a bilayer PbSe unit, forming natural multilayer heterostructure of TI (Bi$_2$Se$_3$) and an ordinary insulator (PbSe) \cite{PSBS1, PSBS2, PSBS3, PSBS4, SegawaPSBS}. An angle-resolved photoemission spectroscopy (ARPES) study has revealed two Dirac-cone states hybridized with each other, suggestive of the existence of topological interface states at the boundaries between 2-QL Bi$_2$Se$_3$ and PbSe layer \cite{NakayamaPSBS}. Upon Cu intercalation into the van der Waals gap at the middle of the 2-QL Bi$_2$Se$_3$ unit, superconductivity with $T_{\rm c}$ $\sim$ 2.9 K has been realized \cite{SasakiCPSBS}. An important distinction of CPSBS from other known TI/TCI-based superconductors is that the unconventional bulk superconductivity with gap nodes is inferred from the specific-heat measurement \cite{SasakiCPSBS}. This is intriguing because the existence of gap nodes in a strongly spin-orbit coupled superconductor would give rise to spin-split Andreev bound states that are nothing but helical Majorana fermions \cite{SasakiCuBi2Se3, NodalTSC}. Since the superconducting pairing symmetry strongly depends on the topology of Fermi surface, it is of particular importance to establish the bulk electronic states relevant to possible unconventional superconductivity in CPSBS.

In this Letter, we report high-resolution ARPES study of PSBS ($m = 2$) and its Cu-intercalated counterpart, CPSBS. By utilizing a relatively long escape depth of photoelectrons excited by low-energy photons, we have succeeded in observing previously unidentified, intrinsic bulk band structure of the 2-QL Bi$_2$Se$_3$ unit lying deeper beneath the surface. Our most important finding is that the near-$E_{\rm F}$ bulk band responsible for the superconductivity shows a two-dimensional (2D) character, in contrast to the 3D character of bulk Bi$_2$Se$_3$. We discuss the implications of the present result in relation to the possible nodal superconductivity and the appearance of Majorana bound states in CPSBS.
 
High-quality single crystals of a parent compound PSBS ($m = 2$) were grown by a modified Bridgman method using high purity elements Pb (99.998 $\%$), Bi (99.9999 $\%$), and Se (99.999 $\%$). The crystal quality has been greatly improved from that in the previous ARPES study of PSBS \cite{NakayamaPSBS}; in particular, we obtained single-phase crystals of $m = 2$ without detectable inclusions of the $m = 1$ phase in the $x$-ray diffraction analysis. Cu intercalation has been achieved by the electrochemical technique, and the superconducting transition temperature ($T_{\rm c}$) of Cu-intercalated sample (CPSBS) was estimated to be $\sim$ 2.9 K by a magnetic susceptibility measurement. The Cu concentration $x$ has been determined to be 1.47 from the mass change and the ICP-AES analyses. Details of the sample preparations were described elsewhere \cite{SasakiCPSBS}. ARPES measurements were performed with MBS-A1 and VG-Scienta SES2002 electron analyzers equipped with a xenon-plasma discharge lamp ($h\nu$ = 8.437 eV) at Tohoku University and also with tunable synchrotron lights of $h\nu$ = 13-23 eV at the beam line BL-7U at UVSOR as well as $h\nu$ = 60 eV at the beam line BL-28A at Photon Factory, respectively. The energy and angular resolutions were set at 10-30 meV and 0.2-0.3$^{\circ}$, respectively. Samples were cleaved {\it in situ} along the (111) crystal plane in an ultrahigh vacuum better than 1 $\times$ 10$^{-10}$ Torr. A shiny mirror-like surface was obtained after cleaving the samples, confirming its high quality. The Fermi level ($E_{\rm F}$) of the samples was referenced to that of a gold film evaporated onto the sample holder.

\begin{figure}
\includegraphics[width=3in]{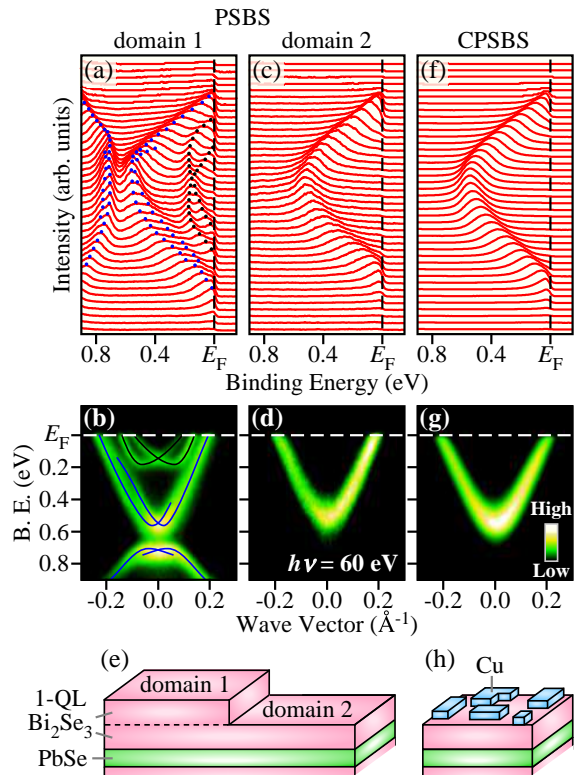}
\vspace{0cm}
\caption{(Color online) (a)-(d) Comparison of near-$E_{\rm F}$ energy distribution curves (EDCs) around the $\bar{\Gamma}$ point [(a) and (c)] and corresponding ARPES intensity [(b) and (d)]. The data in (a),(b) and (c),(d) were taken on different domains of the cleaved surface of pristine (PbSe)$_5$(Bi$_2$Se$_3$)$_6$ (PSBS). All the data have been obtained at $T$ = 30 K with 60-eV photons. Black and blue dots in (a) [as well as black and blue curves in (b)] are a guide to the eyes to trace the dispersions of bulk conduction band and gapped Dirac-cone bands, respectively. (e) Schematic illustration of the cleaved surface of PSBS. (f),(g) Near-$E_{\rm F}$ EDCs and corresponding ARPES intensity plot, respectively, for Cu$_{1.47}$(PbSe)$_5$(Bi$_2$Se$_3$)$_6$ (CPSBS) measured with $h\nu$ = 60 eV. (h) Schematic illustration of the cleaved surface of CPSBS.}
\end{figure}

First, we present high-resolution ARPES data measured with $h\nu$ = 60 eV for pristine PSBS ($m = 2$), where we revealed presence of two domains at the surface (named domain 1 and 2), each of which is selectively probed by scanning the position of a cleaved surface with a small incident beam spot (a diameter of $\sim$100 $\mu$m). Comparison of the energy distribution curves (EDCs) and corresponding ARPES intensity plot around the $\bar{\Gamma}$ point of the Brillouin zone in Figs. 1(a)-1(d) display a marked difference in the near-$E_{\rm F}$ band structure between the two surface domains. For domain 1, the observed band structure resembles the previous ARPES result for pristine PSBS ($m = 2$) whose surface is terminated by a 2-QL Bi$_2$Se$_3$ unit \cite{NakayamaPSBS}. Two inner electron pockets indicated by black dots in Fig. 1(a) and black curves in Fig. 1(b) are ascribed to the quantized bulk conduction band (CB) of 2-QL Bi$_2$Se$_3$ with the Rashba splitting, while a pair of outer-electron and hole bands (blue dots and blue curves) represent hybridized surface and interface Dirac-cone states. On the other hand, the band structure for domain 2 shows a single parabola [Figs. 1(c) and 1(d)], which resembles the quantized CB of the topmost 1-QL Bi$_2$Se$_3$ layer for PSBS ($m = 1$) \cite{NakayamaPSBS}. These results strongly suggest that domain 1 and 2 correspond to 2-QL and 1-QL Bi$_2$Se$_3$ terminations, respectively, as schematically illustrated in Fig. 1(e). Taking into account the single-phase ($m = 2$) nature of our crystal, the different surface terminations are naturally understood by the presence of two cleavage planes for PSBS ($m = 2$), $i.e.$, (i) between PbSe and Bi$_2$Se$_3$ units, and (ii) between 2 QLs within the Bi$_2$Se$_3$ unit. We note that our growth technique for the PSBS system has been greatly improved since our original report \cite{NakayamaPSBS}, and large single-phase crystals for $m = 2$ have become regularly available \cite{SegawaPSBS}.

\begin{figure*}
\includegraphics[width=6.4in]{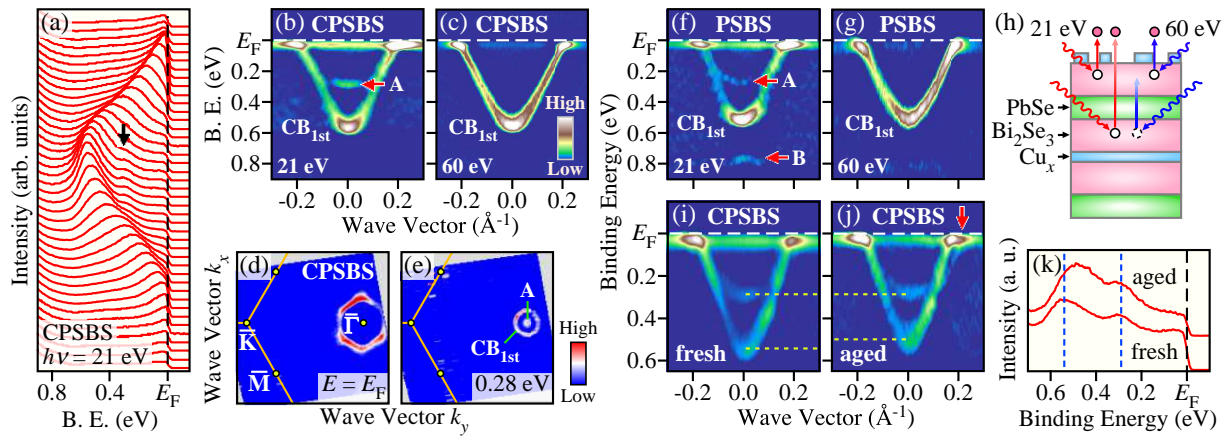}
\vspace{0cm}
\caption{(Color online) (a) EDCs around the $\bar{\Gamma}$ point for CPSBS measured at $T$ = 30 K with $h\nu$ = 21 eV. (b) Second-derivative intensity of (a) plotted as a function of binding energy and wave vector. (c) Same as (b), but measured with $h\nu$ = 60 eV. (d),(e) ARPES intensity map at $E_{\rm F}$ and 0.28-eV binding energy, respectively, for CPSBS plotted as a function of in-plane wave vector. The intensity was obtained by integrating the second-derivative EDCs within $\pm$10 meV centered at $E_{\rm F}$ or 0.28 eV. (f),(g) Second-derivative ARPES intensity around the $\bar{\Gamma}$ point for PSBS at $h\nu$ = 21 and 60 eV, respectively. (h) Schematic illustration of the photoemission process for $h\nu$ = 21 and 60 eV. (i),(j) Near-$E_{\rm F}$ band dispersions obtained soon after cleaving and $\sim$10 hours after cleaving, respectively. (k) Comparison of the EDCs at the $\bar{\Gamma}$ point for fresh and aged surfaces of CPSBS.}
\end{figure*}

Next we have performed the ARPES measurements on the Cu-intercalated counterpart (CPSBS). We found that the observed spectral feature always shows a simple parabolic band [Figs. 1(f) and 1(g)] irrespective of the incident beam position on the cleaved surface, unlike the case of PSBS. As one can immediately recognize from a side by side comparison of Figs. 1(d) and 1(g), the observed band dispersion for CPSBS is very similar to that for domain 2 of PSBS (except for the small downward energy shift due to electron doping). This suggests that the CPSBS sample is always cleaved at the Cu layer as illustrated in Fig. 1(h), presumably due to the expanded van der Waals gap by Cu intercalation. We thus conclude that the observed parabolic band is the quantized CB of the topmost 1-QL Bi$_2$Se$_3$ (hereafter we call CB$_{\rm 1st}$) partially covered with residual Cu atoms on the surface, and it may not reflect genuine bulk electronic properties. Nevertheless, we demonstrate in the following that the utilization of lower-energy photons ($h\nu$ = 8.4-23 eV), which have a relatively long photoelectron escape depth (10-50 {\text \AA}) compared to that for 60-eV photons ($\sim$5 {\text \AA}), allows us to investigate more intrinsic bulk electronic states of the 2-QL Bi$_2$Se$_3$ unit which lies beneath the PbSe layer.

Figures 2(a) and 2(b) display the EDCs of CPSBS and corresponding second-derivative intensity plot, respectively, measured with $h\nu$ = 21 eV. In addition to CB$_{\rm 1st}$, one can find another band at a binding energy of $\sim$ 0.28 eV (labeled A), which was not clearly visible with $h\nu$ = 60 eV [Fig. 2(c)]. This band displays an electronlike dispersion around the $\bar{\Gamma}$ point [see also Fig. 2(i)] and merges to CB$_{\rm 1st}$ with approaching $E_{\rm F}$ (this point will be clarified by the surface aging measurement as shown later). This can be also recognized from the ARPES intensity mappings as a function of 2D wave vector [Figs. 2(d) and 2(e)], where two well-separated circular intensity profiles at 0.28 eV merge to form a hexagonal Fermi surface at $E_{\rm F}$. As shown in Fig. 2(f), we also find a similar electronlike band with $h\nu$ = 21 eV at domain 2 of PSBS together with a holelike band (labeled B), both of which are not seen in the ARPES data for $h\nu$ = 60 eV [Fig. 2(g)]. Since the photoelectron escape depth for $h\nu$ = 21 eV is longer than that for 60 eV as schematically illustrated in Fig. 2(h), it is natural to ascribe bands A and B to the electronic states of the 2-QL Bi$_2$Se$_3$ unit located deeper beneath the surface. In fact, this conclusion is supported by intentionally aging the sample surface. As shown in Figs. 2(i)-2(k), the surface aging leads to a clear upward shift of CB$_{\rm 1st}$ in contrast to the stationary nature of band A, indicating that band A is insensitive to the surface chemical environment as expected for the bulk electronic states. Thanks to the energy shift of CB$_{\rm 1st}$, the $E_{\rm F}$ crossing of band A is now clearly visible as marked by red arrow in Fig. 2(j).

Having established the bulk 2-QL Bi$_2$Se$_3$ origin of bands A and B, a next question is their band assignment. In this regard, it is useful to compare their energy dispersions with those for domain 1 of PSBS, since both originates from the 2-QL Bi$_2$Se$_3$ unit. As shown in Fig. 2(f), the energy separation between the bottom of the electronlike band A and the top of the holelike band B is 0.5 eV. This value is in good agreement with an energy separation between the quantized bulk CB and the lower branch of the gapped Dirac-cone state in the topmost 2-QL Bi$_2$Se$_3$ of PSBS in Fig. 1(a). Thus, bands A and B are attributed to the quantized bulk CB and the lower branch of the gapped Dirac-cone state, respectively. It is noted that bands A and B in Fig. 2(f) are shifted downward by $\sim$ 0.1 eV as compared to those in Fig. 1(a). This may be due to the weakened band-bending effect ($i.e.$ weakened out-of-plane potential gradient) at the second Bi$_2$Se$_3$ unit, which may also explain the absence of a clear Rashaba splitting for band A in Figs. 2(b) and 2(f).

\begin{figure}
\includegraphics[width=3in]{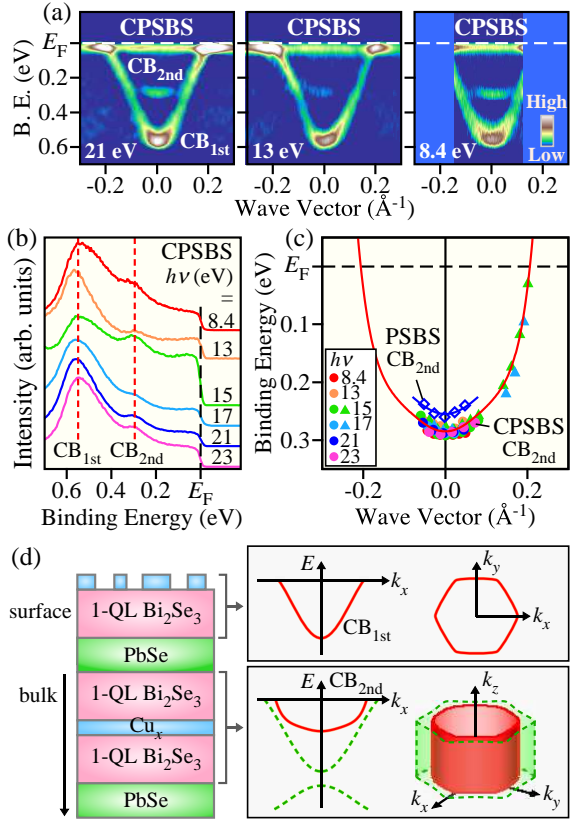}
\vspace{0cm}
\caption{(Color online) (a) Second-derivative ARPES intensity around the $\bar{\Gamma}$ point for CPSBS at various photon energies plotted as a function of binding energy and wave vector. CB$_{\rm 1st}$ and CB$_{\rm 2nd}$ denote the conduction bands stemming from the topmost 1-QL Bi$_2$Se$_3$ unit and the second Bi$_2$Se$_3$ unit located beneath the PbSe layer, respectively. (b) Comparison of the EDC at the $\bar{\Gamma}$ point measured with various photon energies. (c) Photon-energy dependence of the band dispersion for CB$_{\rm 2nd}$ in CPSBS extracted from the second-derivative intensities. Filled circles and triangles represent the results for the fresh and aged surfaces, respectively. Red curve is a guide for the eyes to trace the band dispersion. The band dispersion for PSBS measured with $h\nu$ = 21 eV is also plotted by open diamonds. (d) Schematic illustration of the band dispersion and the Fermi surface in the topmost and the second Bi$_2$Se$_3$ units in CPSBS.}
\end{figure}

To elucidate the Fermi-surface topology in the 3D momentum space relevant to the unconventional superconductivity of CPSBS, we have performed $h\nu$-dependent ARPES measurements by focusing on band A (hereafter we call CB$_{\rm 2nd}$). As one can see in the representative second-derivative intensity plots in Fig. 3(a), we commonly see CB$_{\rm 2nd}$ and CB$_{\rm 1st}$ over a wide $h\nu$ range of 8.4-23 eV (note that 8.4-eV photons are most bulk sensitive). A direct comparison of the EDC at the $\bar{\Gamma}$ point in Fig. 3(b) measured with various photon energies reveals that the energy position of CB$_{\rm 2nd}$ (and also CB$_{\rm 1st}$) is stationary with the $h\nu$ variation in contrast to the 3D-like CB of bulk Bi$_2$Se$_3$ \cite{HasanBi2Se3}. Indeed, the extracted band dispersions of CB$_{\rm 2nd}$ overlap very well with each other within the experimental uncertainty irrespective of the $h\nu$ value [Fig. 3(c)], demonstrating the 2D nature of CB$_{\rm 2nd}$. We also clarified a finite energy shift ($\sim$30 meV) of CB$_{\rm 2nd}$ upon Cu intercalation [see Fig. 3(c)], indicating that the intercalated Cu atoms provide electrons to the Bi$_2$Se$_3$ layers, in line with the Hall-coefficient measurements \cite{SasakiCPSBS}. Based on the present ARPES results, we show in Fig. 3(d) the schematic illustration of electronic structure in CPSBS. While the topmost 1-QL Bi$_2$Se$_3$ produces a single parabolic band (CB$_{\rm 1st}$) localized at the surface, the bulk electronic states are confined within the 2-QL Bi$_2$Se$_3$ unit sandwiched by the insulating PbSe layers and form a cylindrical Fermi surface at the Brillouin-zone center [as shown in red in the right bottom panel of Fig. 3(d)], possibly with hybridized topological interface state (green dashed curves). It is noted though that the topological interface state is not clearly seen in the present experiment, presumably due to its fairly weak intensity or a possible change in the band parity in the 2-QL Bi$_2$Se$_3$ unit upon Cu intercalation.

Now we discuss the implications of our results in relation to the possible nodal superconductivity inferred from a recent specific-heat measurement \cite{SasakiCPSBS}. The most important finding of the present study is that the unconventional superconductivity in CPSBS originates from 2D bulk electronic states. This puts strong constraints on the possible superconducting pairing symmetry, since the pairing is sensitive to the Fermi-surface topology. Specifically, while the previous theoretical studies for Cu$_x$Bi$_2$Se$_3$ with the rhombohedral $D_{3d}$ crystalline symmetry predicted several types of possible nodal pairing states depending on the Fermi-surface shape \cite{FuTSC, SasakiCuBi2Se3, HashimotoGap}, the appearance of gap nodes is restricted to the odd-parity $E_u$ pairing states when 2D Fermi surface is realized \cite{HashimotoGap}. Assuming that the theoretical model proposed for Cu$_x$Bi$_2$Se$_3$ is also applicable to CPSBS, the observed 2D cylindrical Fermi surface would support the odd-parity $E_u$ pairing in CPSBS. It should be noted, however, that the crystal structure of CPSBS has the $C_{2h}$ point group symmetry due to the presence of the PbSe layer with square symmetry, unlike the $D_{3d}$ symmetry of Cu$_x$Bi$_2$Se$_3$. The different crystal symmetry may favor unconventional nodal pairing that was not discussed in Cu$_x$Bi$_2$Se$_3$. Indeed, a recent theoretical study for CPSBS suggested that $d$-wave superconductivity could emerge in the case of 2D Fermi surface \cite{FuPrivateCommun}. In this regard, the present result is compatible with the realization of either the odd-parity $E_u$ pairing or the even-parity $d$-wave pairing. Both states give rise to vertical line nodes in agreement with the temperature dependence of electronic specific heat \cite{SasakiCPSBS}. When vertical line nodes are realized on the 2D Fermi surface of a strongly spin-orbit coupled superconductor, spin-split Andreev bound states, which are the hallmark of topological superconductivity, would emerge on some side surfaces that are perpendicular to the (111) crystal plane. Therefore, the superconducting state of CPSBS may provide an intriguing platform to explore Majorana fermions.

In conclusion, we have reported high-resolution ARPES results on CPSBS to elucidate the band structure relevant to the unconventional superconductivity with possible nodes in the gap function. By utilizing tunable low-energy photons, we have succeeded in separately determining the band dispersion of the 1-QL Bi$_2$Se$_3$ unit at the topmost surface and that of the 2-QL Bi$_2$Se$_3$ unit lying in the bulk. The obtained bulk band structure gives evidence for the formation of a cylindrical Fermi surface in CPSBS. The present result puts strong constraints on the possible pairing symmetry and the location of the emergent Andreev bound states.

\begin{acknowledgements}
We thank T. Toba for his help in the crystal growth. We also thank T. Shoman, H. Kumigashira, K. Ono, M. Matsunami, and S. Kimura for their assistance in the ARPES measurements. This work was supported by JSPS (KAKENHI 23224010, 26287071, 25287079, 24540320, 25220708 and Grant-in-Aid for JSPS Fellows 23.4376), MEXT of Japan (Innovative Area ``Topological Quantum Phenomena"), AFOSR (AOARD 124038), Inamori Foundation, the Murata Science Foundation, KEK-PF (Proposal number: 2012S2-001), and UVSOR (Proposal No. 24-536).
\end{acknowledgements}

\bibliographystyle{prsty}

\end{document}